\documentclass[11pt,twoside]{article}
\usepackage{asp2010}

\resetcounters

\begin{document}

\title{Connections between Quasi-periodicity and Modulation
in Pulsating Stars}
\author{J\'ozsef~M.~Benk\H{o}, and Margit~Papar\'o}
\affil{Konkoly Observatory, MTA CSFK, Konkoly Thege u. 15-17, H-1121 Budapest, Hungary}

\begin{abstract}
The observations of the photometric space telescopes
CoRoT and Kepler show numerous Blazhko RR Lyrae stars
which have non-repetitive modulation cycles.
The phenomenon can be explained by multi-periodic, stochastic
or chaotic processes. From a mathematical point of view, 
almost periodic functions describe all of them. 
We assumed band-limited almost periodic functions 
either for the light curves of the main pulsation or for the
modulation functions. The resulting light curves can generally
be described analytically and it can also be 
examined by numerical simulations. 
This presentation is a part of our systematic study
on the modulation of pulsating stars 
\citep{benko09,benko11,benko12}.
\end{abstract}

\section{Definitions}

When we speak about quasi-periodic signals in 
physics we generally think about signals which can be described 
mathematically by almost periodic functions and not quasi-periodic 
ones. Before defining them let we remind the well-known definition of 
periodic function. $x(t)$ real function is {\it periodic} with 
the period $P$ if
\begin{equation}
x(t)=x(t+P), \ \ {\mathrm {or}} \ \  \vert x(t) - x(t+P) \vert = 0 \ \ 
\ \mathrm{for \ all \ } t.
\end{equation}
Its Fourier representation is
\begin{equation}\label{F0}
x(t)=\frac{a_0}{2} +
\sum_{n=1}^{\infty}{a_n \sin \left( 2\pi n f_0 t + \varphi_n \right)},
\end{equation}
where $f_0$, $a_n$, $\varphi_n$ are constants.

The definition of almost periodic function is quite intuitive. $x(t)$ 
real function is a {\it band-limited almost periodic 
function}\footnote[1]{There are numerous non-equivalent definitions for
almost periodic functions in the literature (see \citealt{bredikhina01}). 
We use now this simple definition.} 
with the period $P$ if
\begin{equation}\label{Qp}
x(t)\approx x(t+P), \ \ {\mathrm {or}} \ \  \vert x(t) - x(t+P) \vert < 
\varepsilon  \ \ \ \mathrm{for \ all \ } t,
\end{equation}
where
\[
0 < \varepsilon \ll \Vert x \Vert = \sqrt {x^2}=
\sqrt{\lim_{\tau\to\infty}\frac{1}{\tau}\int_{-\tau /2 }^{\tau 
/2}{x^2(t)dt}. }
\]
Its Fourier representation is
\begin{equation}\label{F1}
x(t)=\frac{a_0(t)}{2} +
\sum_{n=1}^{\infty}{a_n(t) \sin \left[ 2\pi \int_0^{t} f_n (\tau)d
\tau + \varphi_n(0) \right]}.
\end{equation}
Taking into account Eq.~(\ref{F1}) the instantaneous frequency is
\[
f_n(t)=n f_0 (t) + \frac{1}{2\pi} \varphi_{n}^{\prime}(t),
\]
where $f_0(t)$, $a_n(t)$, $\varphi_n(t)$ are changing in time, but 
slowly. An important consequence of this expression is
that the ``harmonics'' are not exactly harmonic: $f_n(t) \neq nf_0(t)$!

%\section{Base Light Curve for Simulations}

\begin{figure}\label{fig:modell}
   \includegraphics[height=18em]{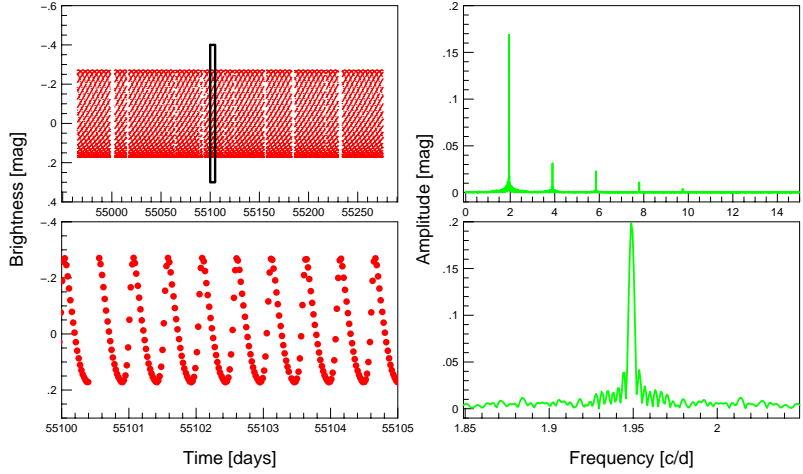}
\caption{
Artificial base light curve used for simulations.
It is prepared from the
Fourier parameters of a typical RRab star and sampled on Kepler 
Q1-Q4 points (top left). (bottom left) A part of the light curve 
(bottom left). Fourier amplitude spectrum of the light 
curve shown in top left panel (top right). 
A zoom of the Fourier spectrum around the main pulsation frequency $f_0$
(bottom right).
}
\end{figure}

For the numerical tests we need a base light curve.
Since this light curve should be
strictly periodic we use the form Eq.~(\ref{F0}) for producing it. 
In practice, we chose Fourier parameters of 
a typical RR Lyrae star's light curve and 
an artificial light curve prepared from them as it is shown in 
Fig.~\ref{fig:modell}.

\section{Quasi-periodic Pulsation}\label{sec:quasi}

\begin{figure}\label{fig:quasi}
   \includegraphics[width=31em]{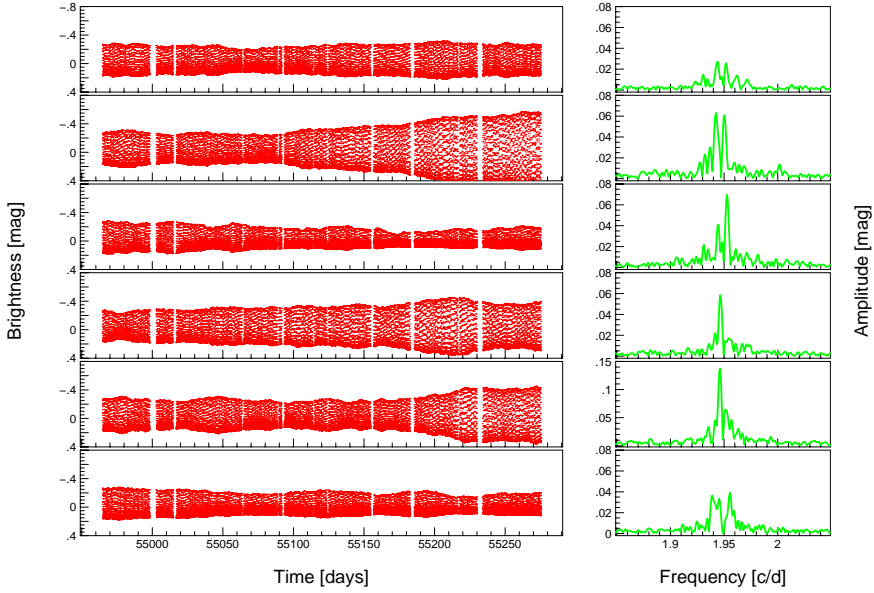}
\caption{ 
Light curves from six different simulation runs. 
The main pulsation frequency and its Fourier
parameters are randomly changed (left panels). 
The Fourier amplitude spectra around the main 
pulsation frequency after 
the data were pre-whitened with it (right panels).
}
\end{figure}

First we made quasi-periodic signals
using the light curve in Fig.~\ref{fig:modell} with the 
assumptions: $f_0(t_{i+1})=f_0(t_i)+k^{\mathrm f} e_i$;
$a_n(t_{i+1})=a_n(t_i)+k^{\mathrm a}_n e_i$ and 
$\varphi_n(t_{i+1})=\varphi_n(t_i)+k^{\varphi}_n e_i$, 
where $e_i$ is the standard white noise process and $k$-s are constants.
The indices $i=1, 2, \dots, $ mean the consecutive sampling points.
In other words, we supposed the time variability to be an
{\it auto-regressive (AR) random process}. This approximation 
tests stochastic signals (see Fig.~\ref{fig:quasi}).
The simulated light curves show correlated 
amplitude and frequency changes, but the 
time scale of the variations support only the longest period 
Blazhko cycles. The corresponding Fourier spectra show
harmonics surrounded by side peaks (even 
multiplets), but generally no peaks can be detected 
in the low frequency regime where the Blazhko frequency 
$f_{\mathrm m}$ can be found for observed stars.
If we change the zero point strongly some peaks
appear in this low frequency range, but we 
parallelly get a strange random walk in the average 
brightness that we never observed in real stars. 

If the quasi-periodicity is caused
by {\it (multi)periodic variation} of which has a characteristic
period(s) $f^{\mathrm m}_j \ll f_0$ for all $j$, we are  
facing a general amplitude and frequency {\it modulated signal}.
Here the modulation functions depend on the harmonics
as it was introduced by \cite{szeidl12}.
It is easy to verify: since  
$a_n(t)$, $f_0(t)$ and $\varphi_n(t)$ functions are periodic,
they can be represented by Fourier series in a form of (\ref{F0}).
Substituting these forms into expression (\ref{F1}) we get
the equations of \cite{szeidl12}. If the modulation function 
always depends on the harmonics of the Blazhko stars 
as stated by \cite{szeidl12}, our formalism suggests 
that even the simplest Blazhko stars have a 
quasi-periodic (at least a multi-periodic) nature. This consequence,
however, needs a careful check and validation. 

\section{Simultaneous Pulsation and Modulation}

As we have seen in the previous section 
the multi-periodicity is in close connection
with the modulation. For simplicity, we investigated the 
modulated signals with global (harmonic independent) modulation 
functions as they were used in \cite{benko11}.
We constructed light curves using the strictly periodic
unmodulated light curve in Fig.~\ref{fig:modell}. 
It is changed with random modulation,
namely, the modulation functions were assumed as
$f^{\mathrm m}(t_{i+1})=f^{\mathrm m}(t_i)+k^{\mathrm f} e_i$
where $e_i$ and $k^{\mathrm f}$ are the white noise process and
a constant term, respectively.
As an alternative we tested cases where the modulation functions
were strictly periodic and the pulsation varied randomly
as it is described in Sec.~\ref{sec:quasi}.

\begin{figure}\label{fig:mod}
\begin{centering}
   \includegraphics[height=23em]{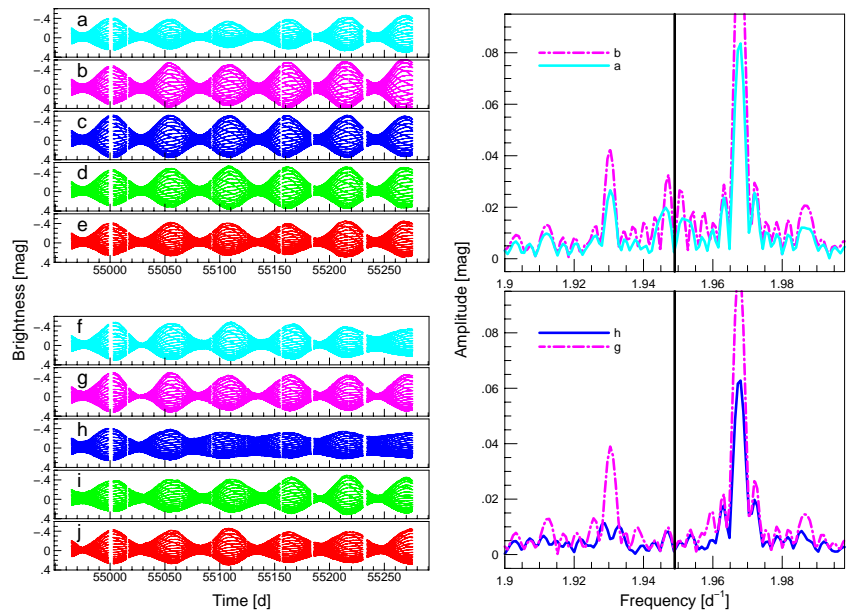}
\end{centering}
\caption{
Light curves from ten simulation runs (left panels).
The five top panels show randomly changing main pulsation 
with fixed modulation, while bottom five ones show fixed pulsation
with randomly perturbed modulation. Fourier spectra in 
left panels are chosen from the spectra of the light curves in right 
panels. The zooms show pre-whitened spectra around the main pulsation 
frequency $f_0$. (Its place is marked by vertical lines in the middle.)
}
\end{figure}

Since the average pulsation and modulation are described
exactly the consecutive runs demonstrate the possible light curve
and Fourier spectra deviations. The simulated light curves 
have very similar characteristics to the observed ones
(see left panels in Fig.~\ref{fig:mod}). They show
changes of the envelope curves, changes in the Fourier amplitudes of the 
modulation frequency and the side peaks;
sometimes additional peaks appear (right panels in Fig.~\ref{fig:mod}).
All of these features are in the same magnitude as the observed effects. 

\section{Conclusions}

We checked two options and their sub-cases: 
almost periodic pulsation alone and 
simultaneous pulsation and modulation where (at least) one 
of them was almost periodic. Random pulsation yields
light curves with amplitude and frequency variation 
but they show significant differences compared to the observed 
light curves. We showed that the strictly periodic sub-case results in a general 
modulation description.

Simultaneous modulation and pulsation where one of them 
has quasi-periodic behaviour yields completely analogous 
light curves and Fourier spectra with the observed data.
The key question is whether the random perturbation
is caused by a stochastic or a chaotic process. 
The best method for determining which one is
the so-called ``phase space reconstruction'' which will be
the next step in our future analysis.

\acknowledgements 
{
The authors acknowledge the support of the 
ESA PECS project No.~4000103541/11/NL/KML.
}

\bibliography{Benko_proc}

\end{document}